# The Use of Quantitative Metrics and Machine Learning to Predict Radiologist Interpretations of MRI Image Quality and Artifacts

**Lucas McCullum**[1,2] (*ORCID: 0000-0001-9788-7987*), **John Wood**[2] (*ORCID: 0000-0002-2040-7735*), **Maria Gule-Monroe**[2] (*ORCID: 0000-0002-4418-3705*), **Ho-Ling Anthony Liu**[2] (*ORCID: 0000-0002-0284-5889*), **Melissa Chen**[2] (*ORCID: 0000-0002-3274-2653*), **Komal Shah**[2] (*ORCID: 0000-0003-2122-9413*), **Noah Nathan Chasen**[2] (*ORCID: 0000-0001-6882-6739*), **Vinodh Kumar**[2] (*ORCID: 0000-0002-8322-1233*), **Ping Hou**[2] (*ORCID: 0000-0002-1518-571X*), **Jason Stafford**[2] (*ORCID: 0000-0003-4091-8417*), **Caroline Chung**[2] (*ORCID: 0000-0002-9662-1519*), **Moiz Ahmad**[2] (*ORCID: 0000-0002-6912-6029*), **Christopher Walker**[2] (*ORCID: 0000-0002-6381-8737*), **Joshua Yung**[2] (*ORCID: 0000-0003-3752-5997*)

[1]The University of Texas MD Anderson Cancer Center UTHealth Houston Graduate School of Biomedical Sciences, [2]The University of Texas MD Anderson Cancer Center

## Abstract

A dataset of 3D-GRE and 3D-TSE brain 3T post contrast T1-weighted images as part of a quality improvement project were collected and shown to five neuro-radiologists who evaluated each sequence for both image quality and imaging artifacts. The same scans were processed using the MRQy tool for objective, quantitative image quality metrics. Using the combined radiologist and quantitative metrics dataset, a decision tree classifier with a bagging ensemble approach was trained to predict radiologist assessment using the quantitative metrics. A machine learning model was developed for the following three tasks: (1) determine the best model / performance for each MRI sequence and evaluation metric, (2) determine the best model / performance across all MRI sequences for each evaluation metric, and (3) determine the best general model / performance across all MRI sequences and evaluations. Model performance for imaging artifact was slightly higher than image quality, for example, the final generalized model AUROC for image quality was 0.77 (0.41 – 0.84, 95% CI) while imaging artifact was 0.78 (0.60 – 0.93, 95% CI). Further, it was noted that the generalized model performed slightly better than the individual models (AUROC 0.69 for 3D-GRE image quality, for example), indicating the value in comprehensive training data for these applications. The use of classification models utilizing quantitative metrics to predict radiologist interpretations has the potential to expedite the quality control and quality assurance protocols currently used today and ensure sufficient image quality required for accurate diagnosis and decision-support. These models could be deployed in the clinic as automatic checks for real-time image acquisition to prevent patient re-scanning requiring another appointment after retrospective radiologist analysis or improve reader confidence in the study. Further work needs to be done to validate the model described here on an external dataset. The results presented here suggest that MRQy could be used as a foundation for quantitative metrics as a surrogate for radiologist assessment.

## Introduction

Magnetic Resonance Imaging (MRI) has become a popular option for patients with 34.7 million receiving an MRI scan in 2022 alone in the United States[1]. However, recent studies have suggested that 7% of outpatient and 29% of inpatient MRI scans exhibited motion artifacts resulting in 20% of all MRI examinations requiring a repeat acquisition costing the hospital, on average, upwards of $100,000 in lost revenue per year[2]. Some MRI sequences or acquisition parameters may be the primary cause of necessitating repeated examinations and, therefore, should be identified by the institution and addressed. Unfortunately, expert-annotated radiological data for image quality assessment and artifact screening is a time-consuming, expensive, and a potentially unreliable quality assessment method due to the presence of, sometimes significant, inter-reader variability (differences in assessment of the same image across annotators). For this reason, alternative methods such as quantitative approaches to measure image quality and artifacts are currently being investigated to reduce the workload of radiologists and increase clinical workflow efficiency and cost-effectiveness[3,4]. The aim of this study was to develop and validate a classification model to automatically predict image quality and level of artifact in MRI images from a retrospective cohort of patients scanned at our institution.

## Materials and Methods

A series of 188 MRI examinations (5% 1.5T, 95% 3T) were acquired solely from our institution from 09/2021 to 02/2022 as part of a Practice Quality Improvement (PQI) project with the intention to compare the application of 3D Gradient Recalled Echo (3D-GRE) and 3D Turbo Spin Echo (3D-TSE) in the brain for detection of metastatic lesions. Patients were eligible if they received either a 3D-GRE or 3D-TSE MRI examination which showed confirmed metastatic lesions. The primary sites of the metastatic lesions were lung (72), breast (41), melanoma (31), renal (18), colon (2), other (14), and not available (10). Five board certified neuroradiologists (M.G.M., M.C., K.S., N.C., and V.K.) worked independently in a blind assessment to evaluate the MRI images using a Likert score for image quality defined as signal-noise-ratio (SNR), noise, contrast, and resolution (1 - unacceptable, 2 - poor, 3 - acceptable, 4 - good, and 5 - excellent) and imaging artifact defined as image degradation by artifact (1 - severe, 2 - moderate, 3 - minor, and 4 - no artifact). A subset (n = 42) of these MRI images were reviewed by two neuroradiologists to evaluate inter-reader variability in addition to Cohen's kappa statistic. Additional information was also collected for each image such as apparent cause of imaging artifact, if present. Figure 1 shows a comprehensive outline of the process used in this study to clean the initial cohort while also providing detailed information on the total number of assessments and duplicate readings for inter-reader variability at each stage. The final number of validated radiologist assessments of image quality were 87 and 86 for the 3D-GRE and 3D-TSE, respectively, and 91 and 88 for the 3D-GRE and 3D-TSE for imaging artifact, respectively.

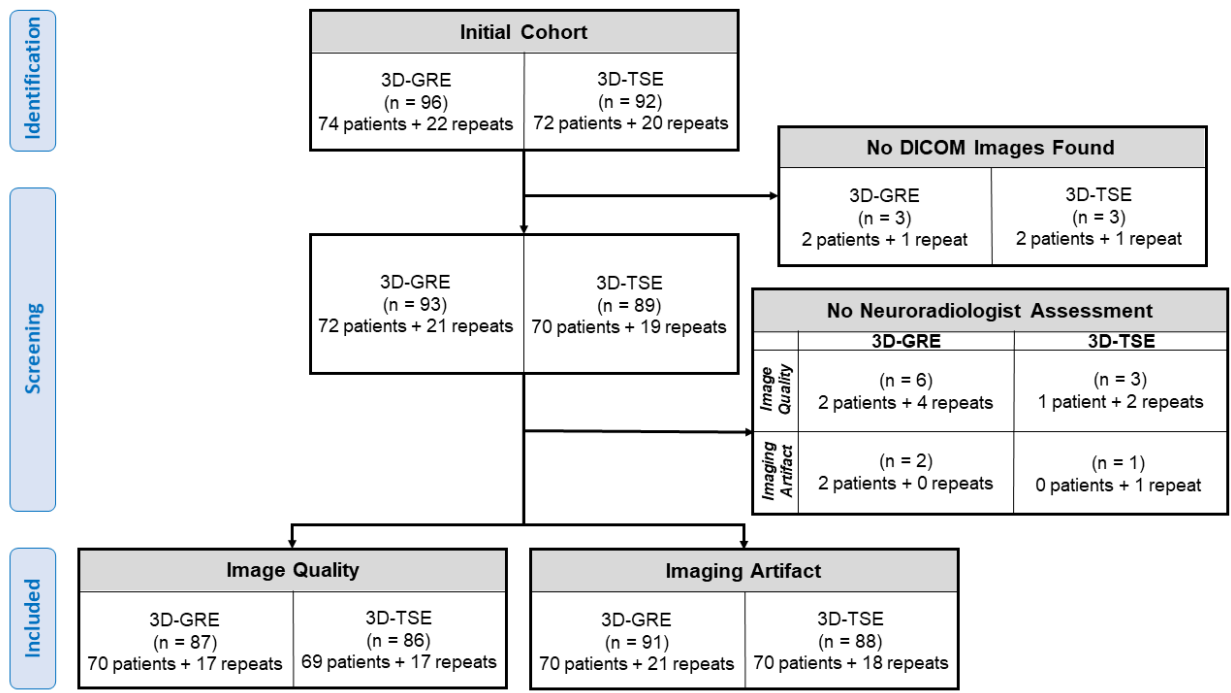

Figure 1: Process of selecting the final patient cohort. Only a small number of images were lost due to either missing DICOM files or having unmatched neuroradiologist assessments. After

filtering, the ratio of number of repeated readings on respective MRI sequence images to the different evaluation metrics stayed relatively constant.

To predict radiologist classifications of image quality and imaging artifact in MRI images based on quantitative metrics, the quality assurance and checking tool, MRQy, was processed on each scan to produce the 25 quantitative metrics described in the tool's original publication[5]. After quantitative imaging metric feature extraction with MRQy, colinear metrics were dropped using a Pearson correlation matrix where highly correlated metrics ($|r| > 0.7$) were reduced to a single metric to prevent overfitting and provide the most robust and generalizable predictions. The decision on which of the two highly correlated metrics to keep was made by manually selecting the most interpretable feature (i.e., coefficient of variation, CV, over the coefficient of joint variation, CJV) while ensuring no further high correlations with the remaining features.

A machine learning model was developed for the following three tasks: (1) determine the best model / performance for each MRI sequence and evaluation metric, (2) determine the best model / performance across all MRI sequences for each evaluation metric, and (3) determine the best general model / performance across all MRI sequences and evaluations. Using this approach, the adoption of the model in a clinical setting may be evaluated with the intention of addressing the following question: For this task, can I employ one generalized model or do I have to fine-tune multiple models dedicated for their respective task? Reporting of the proposed machine learning predictive model followed the Transparent Reporting of a multivariable prediction model for Individual Prognosis or Diagnosis (TRIPOD) guidelines[6].

A decision tree classifier was employed and validated over a series of hyperparameters including maximum tree depth, minimum samples to split, and minimum samples per leaf with constant parameters such as Gini Impurity criterion and seed number for reproducibility of random sampling. The input to the model was split into 70% training and 30% testing subsets and to minimize overfitting and generalize over many subsets of the training data, a bagging classifier[7] with 1000 estimators was applied on the original decision tree classifier. Secondly, reduction of the majority class was applied by ensuring an equal number of samples with the minority class and evaluated for all cases except for 3D-TSE imaging artifact due to an insufficient number of samples for cross validation (50% reduction was used instead). For the purposes of this study, to flag which images to reject and accept a binary cutoff was established for image quality (1-3 reject, 4-5 accept) and imaging artifact (1-2 reject, 3-4 accept).

The model output was tested using both a Receiver Operator Characteristic (ROC) Curve and Precision-Recall (PR) Curve with subsequent Area Under the ROC Curve (AUROC) and

Area Under the PR Curve (AUPRC) analysis, respectively. The 95% confidence intervals were determined using 100 repetitions of stratified 5-fold cross validation to ensure sufficient sampling size and class balance for each fold. Gini importance was determined for each model to evaluate the most important features while the F-score with β=2 (F2-score) was calculated to help evaluate which model performed best at precision-based tasks where the positive class was the reject label. An overview of the study and subsequent analysis is shown in Figure 2.

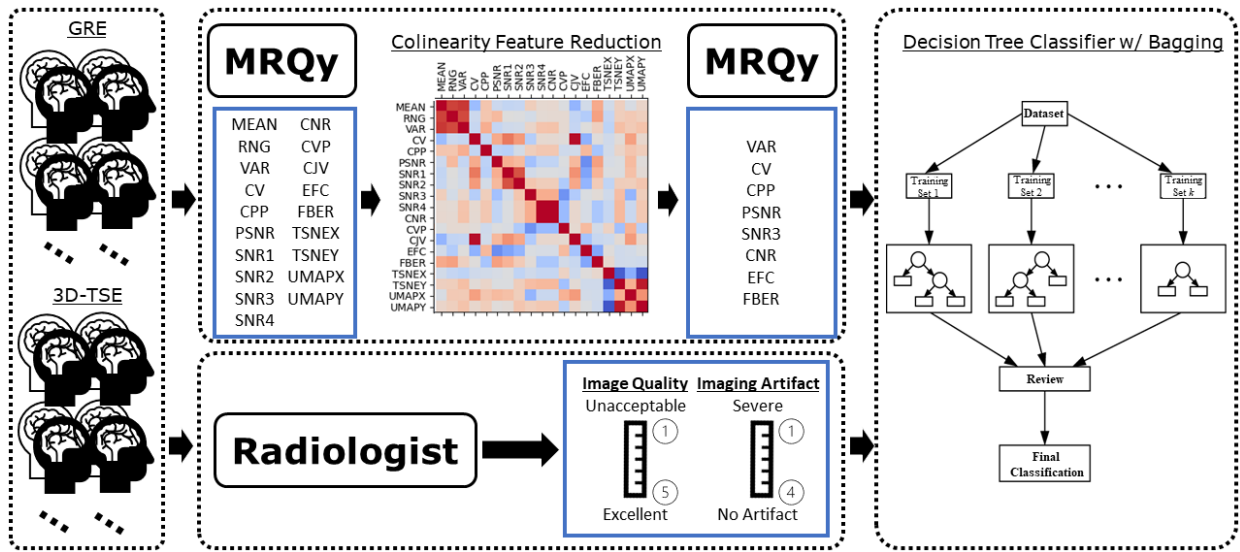

Figure 2: A simplified visual abstract of the project methods. A collection of 3D-GRE and 3D-TSE brain scans were acquired for a set of patients. These scans were interpreted by a group of neuroradiologists providing image quality and artifact scores on a Likert scale. In parallel, the MRQy tool was run for each scan providing quantitative metrics which were reduced after using the Pearson correlation matrix. The combined quantitative imaging metrics (i.e., the predictor variables) and radiologist scores (i.e., the labels) results were used to train and test a decision tree classifier with bagging ensemble methods.

## Results

The outcome of the extensive neuroradiologist evaluation of MRI images for two separate sequences, 3D-GRE and 3D-TSE for image quality and imaging artifact are shown in Table 1.

| *Image Quality* | 3D-GRE (n = 87) | 3D-TSE (n = 86) | Total (n = 173) |
| --- | --- | --- | --- |

| Image Quality | 3D-GRE (n = 87) | 3D-TSE (n = 86) | Total (n = 173) |
|---|---|---|---|
| 1 – Unacceptable | 0 | 0 | 0 |
| 2 – Poor | 5 | 1 | 6 |
| 3 – Acceptable | 22 | 10 | 32 |
| 4 – Minor | 39 | 26 | 65 |
| 5 – None | 21 | 49 | 70 |
| *Imaging Artifact* | *3D-GRE (n = 91)* | *3D-TSE (n = 88)* | *Total (n = 179)* |
| 1 – Severe (non-diagnostic) | 1 | 0 | 1 |
| 2 – Moderate (non-diagnostic) | 23 | 3 | 26 |
| 3 – Minor (diagnostic) | 52 | 37 | 89 |
| 4 – None (diagnostic) | 15 | 48 | 63 |

Table 1: Distribution of radiologist assessments of image quality and imaging artifact for the 3D-GRE and 3D-TSE MRI sequences for the final patient cohort. A majority of scans were either no defects or only minor defects with 3D-GRE displaying worse metrics on average compared to 3D-TSE for both image quality and imaging artifact. The most common causes of imaging artifact identified by the neuroradiologists were flow artifacts (36%), motion (30%), noise (24%), other (1.6%), and susceptibility (0.5%).

When evaluating inter-reader variability for the data shown in Table 1, Cohen's kappa statistic of 0.12, 0.10, 0.03, and 0.10 was found for the neuroradiologist assessment of 3D-GRE image quality, imaging artifact, 3D-TSE image quality, and imaging artifact, respectively. In addition, it was found that 47%, 52%, 41%, and 56% of reviewers agreed on the same label for 3D-GRE image quality, imaging artifact, 3D-TSE image quality, and imaging artifact, respectively with two conflicts between label 5 and 3 for 3D-GRE image quality, one conflict between label 4 and 2 for 3D-GRE imaging artifact, and two conflicts between label 5 and 3 for 3D-TSE image quality.

| | *Maximum Tree Depth* | *Minimum Samples to Split* | *Minimum Samples per Leaf* |
|---|---|---|---|
| 3D-GRE Image Quality | 2 | 8 | 3 |
| 3D-GRE Imaging Artifact | 4 | 2 | 1 |
| 3D-TSE Image Quality | 2 | 2 | 3 |
| 3D-TSE Imaging Artifact | 2 | 2 | 1 |
| 3D-GRE /3D-TSE Image Quality | 2 | 2 | 11 |
| 3D-GRE /3D-TSE Imaging Artifact* | 2 | 2 | 1 |
| General Model* | 2 | 2 | 1 |

Table 2: The final architectures for each optimal model displaying only the three decision tree classifier hyperparameters which were adjusted for this study. Almost all of the optimal models have a maximum tree depth of 2 to prevent overfitting and a minimum samples to split also of 2 for simplicity. The minimum samples per leaf of 1 is typically default, however sometimes it needs to be adjusted higher to prevent overfitting as well. *The same optimal model was found for both of these cases.

After filtering, the remaining nine MRQy metrics used to make the classifications were variance of the foreground (VAR), coefficient of variation (CV), contrast per pixel (CPP), peak signal to noise ratio (PSNR), foreground patch standard deviation divided by the centered foreground patch standard deviation (SNR3), contrast to noise ratio (CNR), coefficient of variation (CVP), entropy focus criterion (EFC), and foreground-background energy ratio (FBER). A detailed summary of the feature weights for each model is shown in Table 3.

| Feature | Feature Weights for Each Model | | | | | | | |
|---|---|---|---|---|---|---|---|---|
| | 3D-GRE Image Quality | 3D-GRE Imaging Artifact | 3D-TSE Image Quality | 3D-TSE Imaging Artifact | 3D-GRE / 3D-TSE Image Quality | 3D-GRE / 3D-TSE Imaging Artifact* | General Model (Image Quality) | General Model (Imaging Artifact)* |
| CNR | 0.05 | 0.07 | 0.18 | 0.01 | 0.02 | 0.10 | 0.06 | 0.10 |
| CPP | 0.13 | 0.11 | 0.07 | 0.04 | 0.01 | 0.08 | 0.06 | 0.08 |
| CV | 0.26 | 0.03 | 0.07 | 0.02 | 0.03 | 0.03 | 0.11 | 0.03 |
| CVP | 0.08 | 0.05 | 0.35 | 0.02 | 0.01 | 0.06 | 0.03 | 0.06 |
| EFC | 0.02 | 0.07 | 0.05 | 0.12 | 0.09 | 0.10 | 0.16 | 0.10 |
| FBER | 0.19 | 0.38 | 0.04 | 0.50 | 0.55 | 0.36 | 0.30 | 0.36 |
| PSNR | 0.04 | 0.09 | 0.09 | 0.11 | 0.04 | 0.13 | 0.08 | 0.13 |
| SNR3 | 0.04 | 0.08 | 0.06 | 0.02 | 0.04 | 0.02 | 0.10 | 0.02 |
| VAR | 0.18 | 0.13 | 0.09 | 0.02 | 0.20 | 0.11 | 0.10 | 0.11 |

Table 3: The feature weights extracted by Gini importance for each model. The most influential features across all models for predicting radiologist interpretations of image quality were FBER, CV, VAR, and SNR3 while the most influential for imaging artifact were FBER, PSNR, VAR, EFC, and CNR. *The same optimal model was found for both of these cases.

The final model performance results are shown in Table 4. Since this is a binary classification class and predicting the reject class is more clinically relevant than predicting the accept class, only the AUROC for the reject class was reported. It should be noted that the AUROC would be the same for the accept class, but the ROC curve would be symmetrically flipped across the line intersecting the top left to the bottom right side of the ROC graph. In addition, the AUPRC for both the reject and accept classes were reported since this may not be symmetric unlike the AUROC.

| | *AUROC Reject (95% C.I.)* | *AUPRC Reject (95% C.I.)* | *AUPRC Accept (95% C.I.)* |
|---|---|---|---|
| 3D-GRE Image Quality | 0.69 (0.57 – 0.77) | 0.77 (0.52 – 0.81) | 0.72 (0.53 – 0.76) |
| 3D-GRE Imaging Artifact | 0.71 (0.64 – 0.84) | 0.80 (0.74 – 0.86) | 0.55 (0.51 – 0.86) |
| 3D-TSE Image Quality | 0.83 (0.17 – 1.00) | 0.91 (0.41 – 1.00) | 0.76 (0.25 – 1.00) |
| 3D-TSE Imaging Artifact | 1.00 (1.00 – 1.00) | 1.00 (0.06 – 1.00) | 1.00 (0.88 – 1.00) |
| 3D-GRE /3D-TSE Image Quality | 0.76 (0.70 – 0.78) | 0.77 (0.72 – 0.80) | 0.80 (0.58 – 0.85) |
| 3D-GRE /3D-TSE Imaging Artifact* | 0.78 (0.60 – 0.93) | 0.83 (0.61 – 0.92) | 0.74 (0.59 – 0.96) |
| General Model (Image Quality) | 0.77 (0.41 – 0.84) | 0.81 (0.46 – 0.85) | 0.78 (0.41 – 0.87) |
| General Model (Imaging Artifact)* | 0.78 (0.60 – 0.93) | 0.83 (0.61 – 0.92) | 0.74 (0.59 – 0.96) |

Table 4: Resulting performance of the models. The results for the 3D-TSE imaging artifact are expected due to massive class imbalance where the test set is only one sample. Otherwise, no major difference is seen between the model trained on only sequence specific data and the generalized model across both sequence and evaluation metrics. *The same optimal model was found for both of these cases.

## Discussion

Although this study only looked at simple machine learning models (e.g., decision trees), it provided encouraging results which motivate future studies to consider neural networks or deep learning methods which could provide better results than those shown here due to their increased parameters and feature filters. Additionally, the low Cohen's kappa statistics (all less than 0.15), with only half of raters agreeing on the same label on average, could have negatively affected the model by providing uncertainty regarding the true labels on some of the images' quality and level of artifact. These results were not fully expected from the authors and further work should be done to investigate the reasons for the high inter-reader variability and how this could affect clinical diagnoses. It should be noted though, that a majority of the discrepancies were in adjacent labels which was another incentive to binarize the labels into acceptable and unacceptable cutoffs to minimize artifacts of study design.

As shown in this study, three of the top five quantitative imaging features for predicting imaging artifact, namely CNR[8], EFC[9], and FBER[10], have also been shown in the past to be sensitive to various artifacts providing validation for the results shown here. After merging the acceptable and unacceptable classes, the 3D-TSE imaging artifact only had 3/88 (3.4%) of available scans in the unacceptable class which caused a high class imbalance leading to lack of trust for the results of that specific study especially considering an AUROC and AUPRC for both evaluations of 1.00. This also demonstrates that the 3D-TSE sequence less frequently causes artifacts when compared to the 3D-GRE sequence so more data should be collected to accurately build this model. With a majority of the AUROCs computed in this study being at or above 0.7, there is promising evidence that the MRQy metrics could be used for rapid estimates of radiologist assessments for both image quality and imaging artifact for both the 3D-GRE and 3D-TSE sequences. Further inspection shows that the generalized models perform slightly better than what would be possible by combining either the four 3D-GRE / image quality, 3D-GRE / imaging artifact, 3D-TSE / image quality, and 3D-TSE / imaging artifact or the combined sequence image quality and combined sequence imaging artifact performances.

In conclusion, we have shown that quantitative metrics of imaging features in MRI images have potential use in automatically identifying images with poor image quality and / or unacceptable imaging artifact. Further, we have shown that generalized models trained on data from two different MRI sequences to evaluate image quality and imaging artifact outperform specific models built for each task, i.e., one model for 3D-GRE scans to predict image quality and another for 3D-TSE scans to predict imaging artifact, etc. These results support further efforts in applying this methodology, the use of quantitative imaging metrics for automated quality control, to reduce radiologist workload and burden while improving quality of diagnostic tools within an institution. With that in mind, broader implications of this work apply what has been shown here to develop system-wide quality control tools so institutions can evaluate not only which patients may require a re-scan, but also which scanning protocols are most likely to elicit this. As a result, institutions can become more responsive to new protocols which underperform existing protocols and make changes where needed by applying this model to predict which scans to reject.

## Conclusion

This study investigated the use of quantitative imaging metrics derived from MRI examinations of the brain using two different sequences to predicted radiologist assessments of image quality and imaging artifact. The MRI examinations were reviewed by at least one radiologist with a subset being reviewed by two radiologists to determine inter-reader variability which was low after Cohen's kappa testing. A decision tree classifier was chosen and fine-tuned to the most optimal hyperparameters for seven different tasks representing each combination of possibilities for a specialized (i.e., predicting image quality of 3D-GRE) to generalized approach (i.e., predicting both image quality and imaging artifact from both 3D-GRE and 3D-TSE). The performance for each model was evaluated and determined to be similar across all tested tasks even performing slightly higher for the more generalized tasks. This data suggests that curated quantitative imaging metrics such as the MRQy metrics could be used for automated image quality analysis and artifact detection to enhance institution's quality controls procedures.

# Acknowledgements

This work was supported by the Tumor Measurement Initiative (TMI) through the MD Anderson STrategic Initiative DEvelopment Program (STRIDE).

# Supplementary

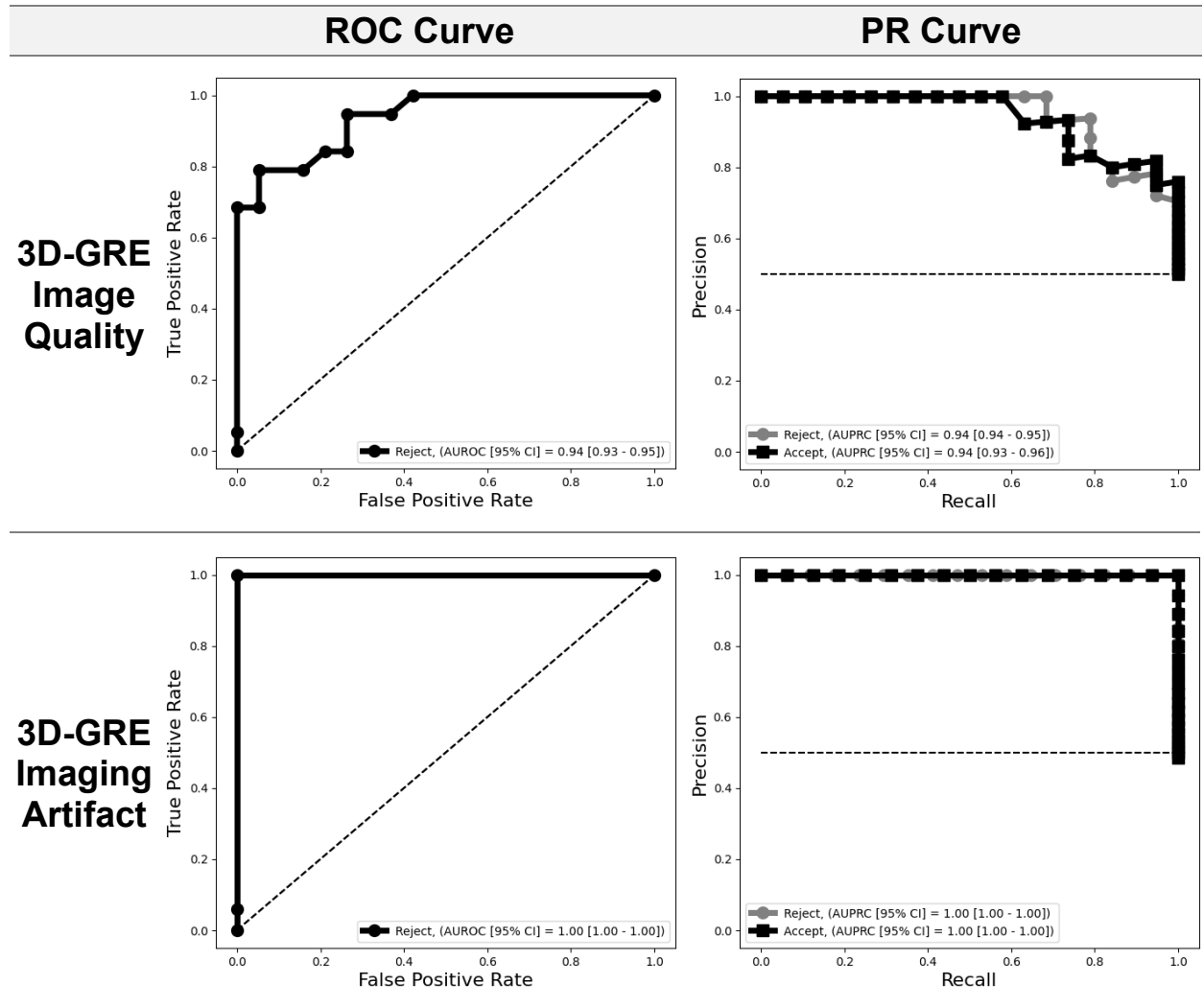

| | | |
|---|---|---|
| **3D-TSE Image Quality** | 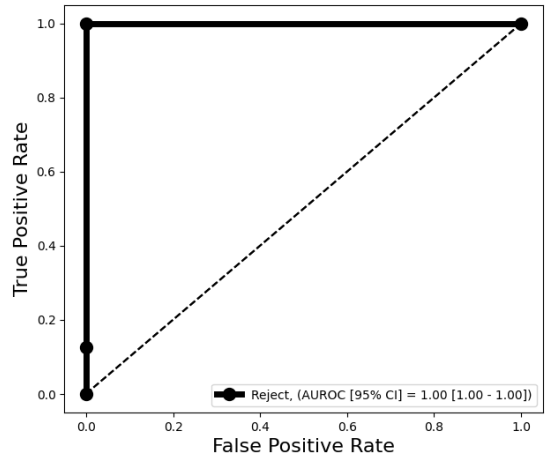 | 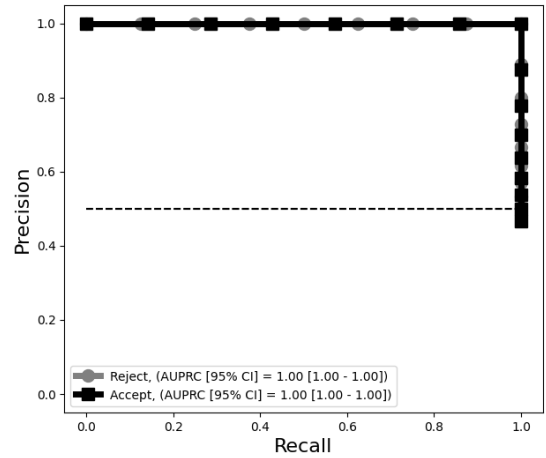 |
| **3D-TSE Imaging Artifact** | 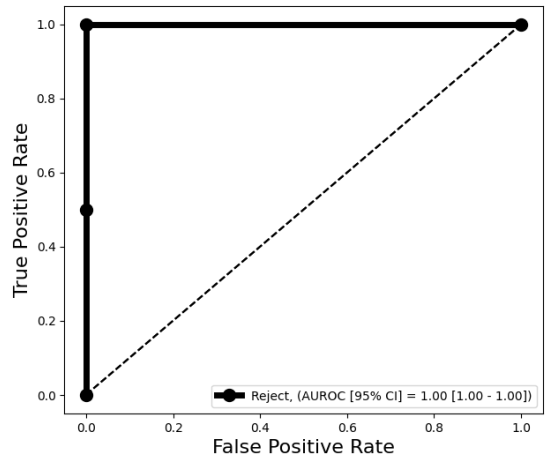 | 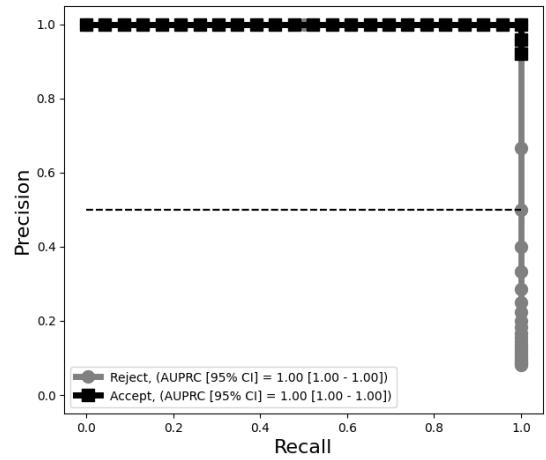 |
| **3D-GRE /3D-TSE Image Quality** | 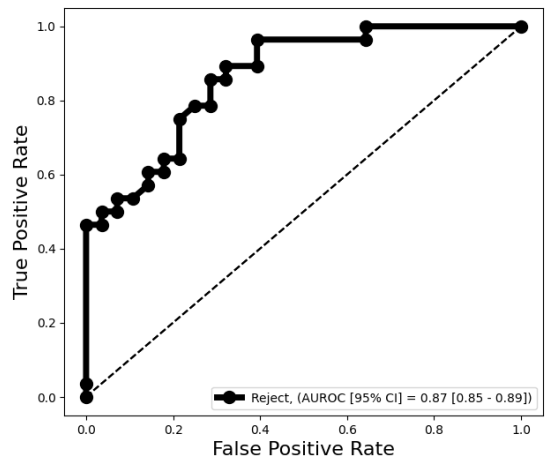 | 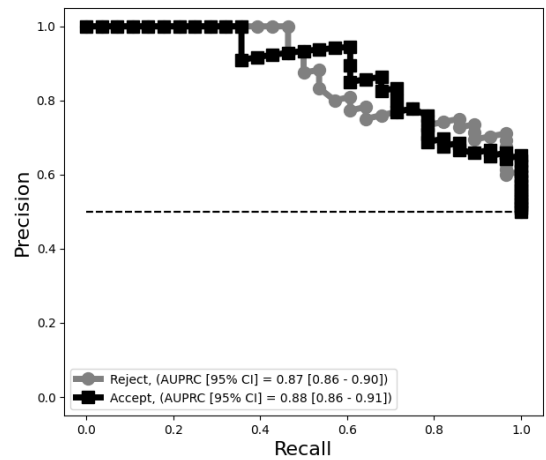 |

**3D-GRE /3D-TSE Imaging Artifact**

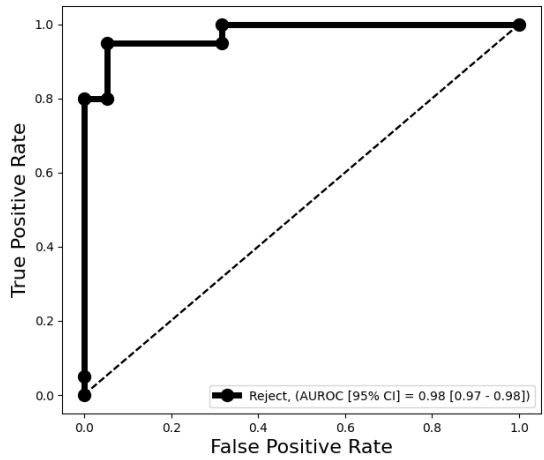
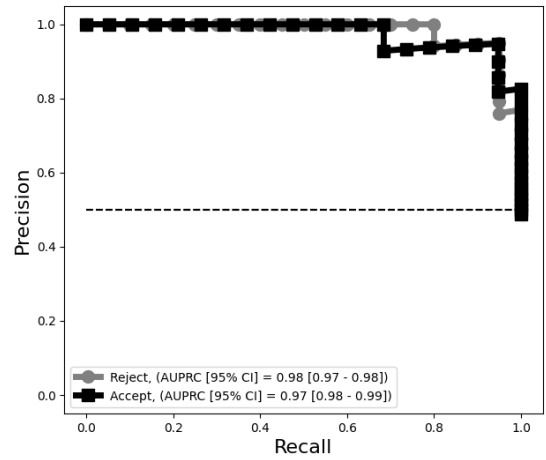

**General Model (Image Quality)**

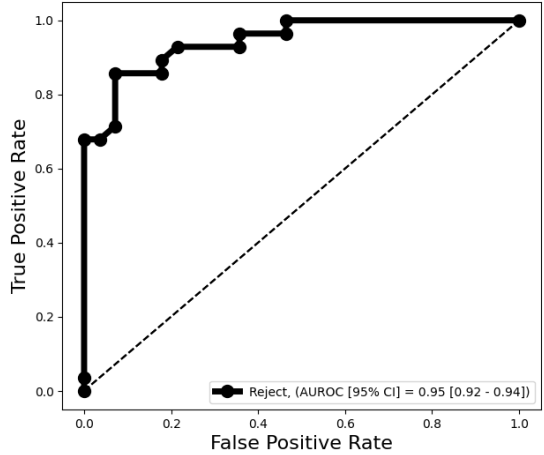
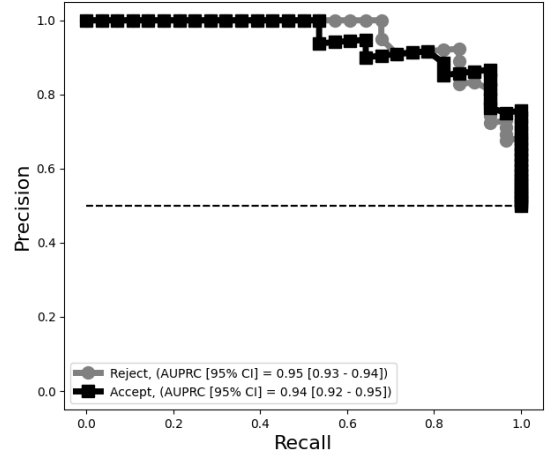

**General Model (Imaging Artifact)**

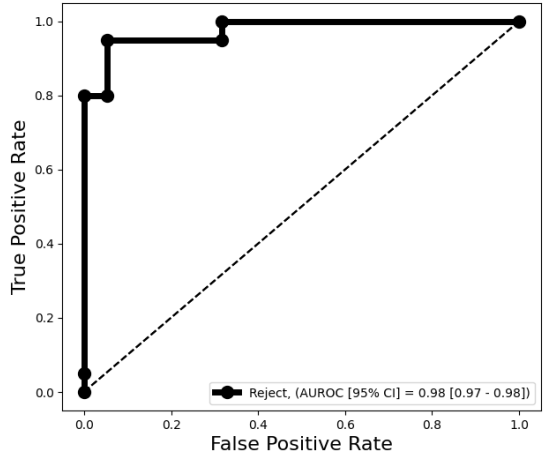
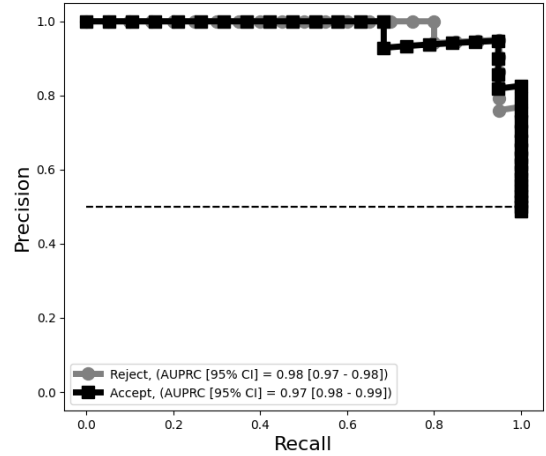

Table S1: Resulting performance of the model on the train set.

|  | ROC Curve | PR Curve |
|---|---|---|
| **3D-GRE Image Quality** | Reject, (AUROC [95% CI] = 0.69 [0.57 - 0.77]) | Reject, (AUPRC [95% CI] = 0.77 [0.52 - 0.81])<br>Accept, (AUPRC [95% CI] = 0.72 [0.53 - 0.76]) |
| **3D-GRE Imaging Artifact** | Reject, (AUROC [95% CI] = 0.71 [0.64 - 0.84]) | Reject, (AUPRC [95% CI] = 0.80 [0.74 - 0.86])<br>Accept, (AUPRC [95% CI] = 0.55 [0.51 - 0.86]) |

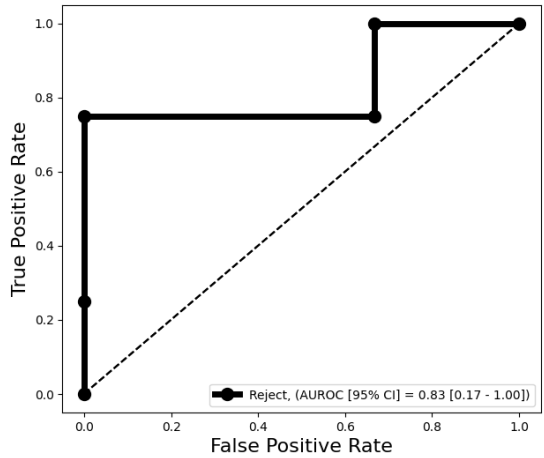
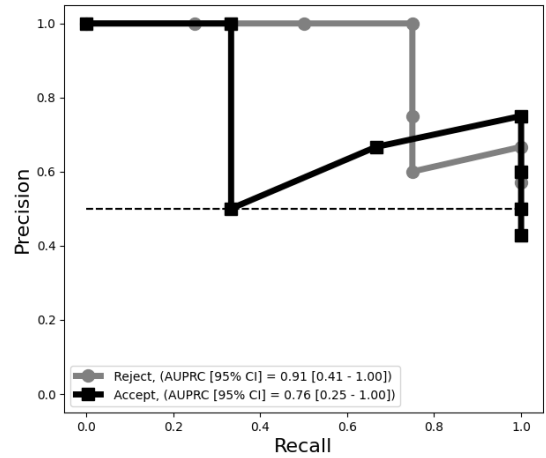
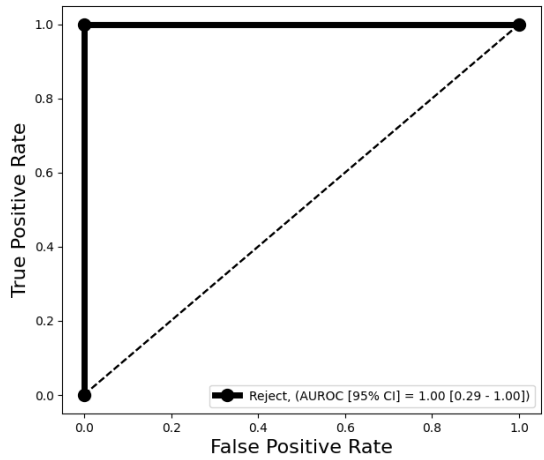
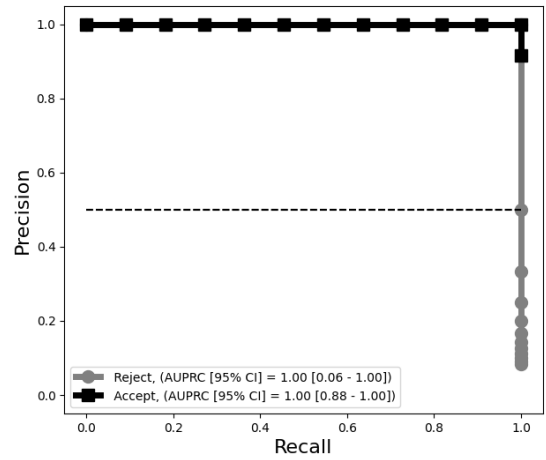
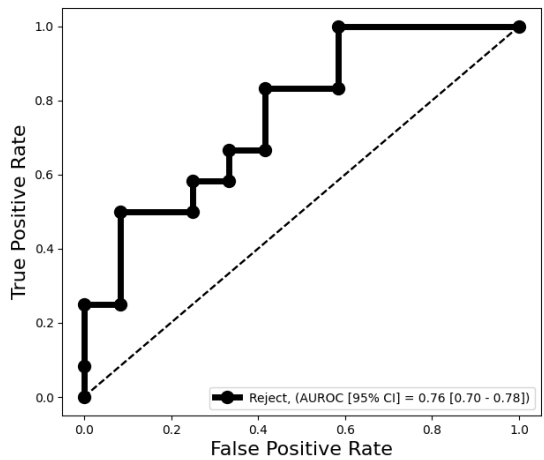
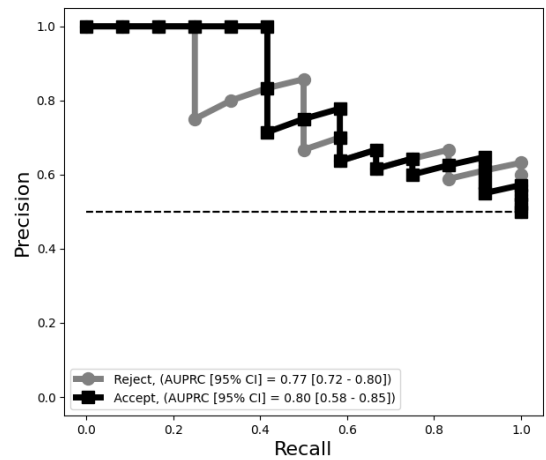

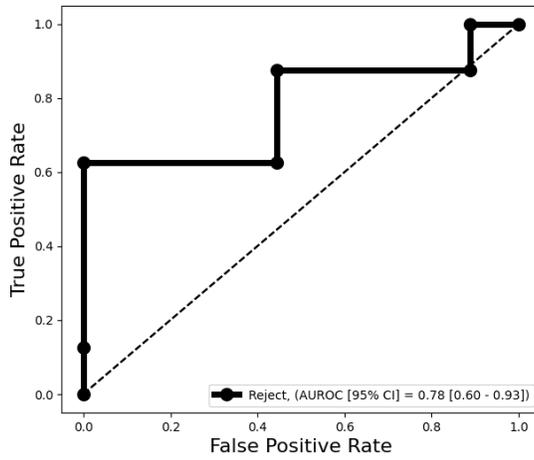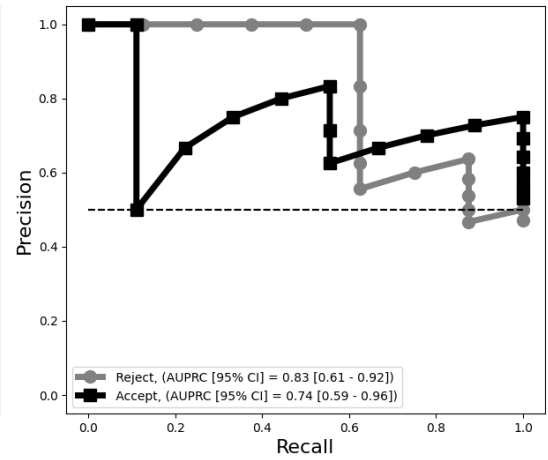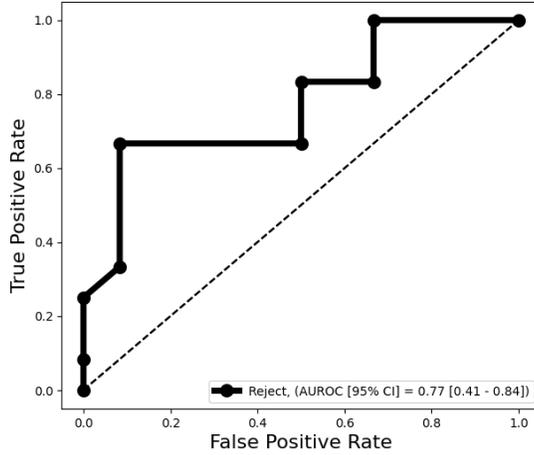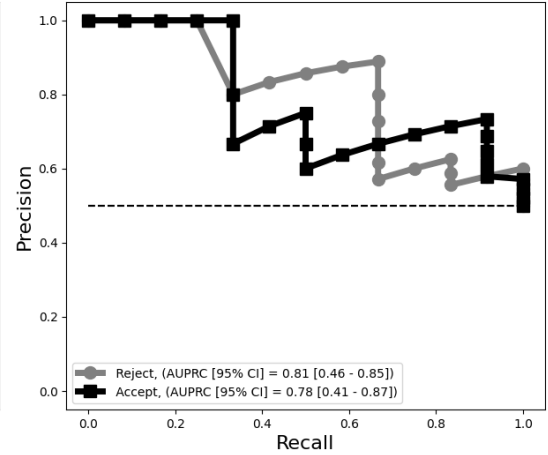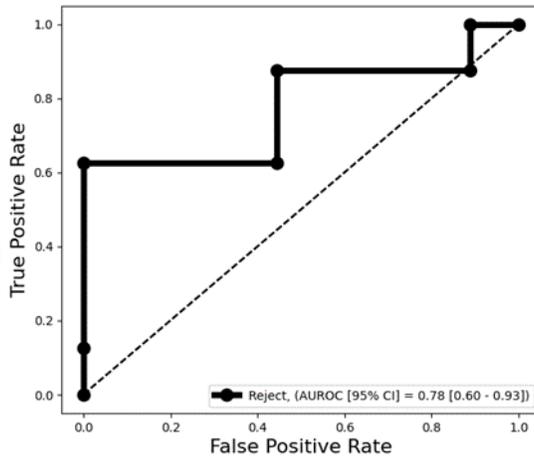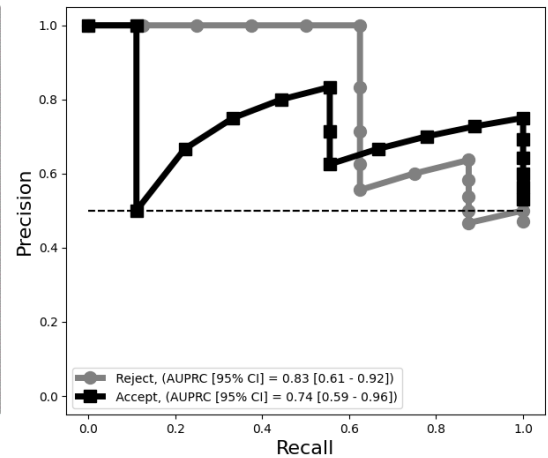

Figure S2: Resulting performance of the model on the test set.